\documentclass[%
showpacs,
prd,
amsmath,amssymb,
article,
nofootinbib
]{revtex4}

\usepackage[pdftex]{graphicx}
\usepackage{bm}
\usepackage{float,color}
\usepackage{comment}
\usepackage{subfig}
\usepackage{bbold}

\newcommand{\eq}[1]{\begin{align} #1 \end{align}}

\newcommand{\bs}[1]{\boldsymbol{#1}}
\newcommand{\pa}{\partial}
\newcommand{\pfrac}[2]{\dfrac{\delta #1}{\delta #2}}

\newcommand{\mH}{\mathcal{H}}
\newcommand{\mM}{\mathcal{M}}

\begin{document}

\title{The Functional Schr\"odinger Equation in the Semiclassical Limit of Quantum Gravity with a Gaussian Clock Field}
\author{Rotondo Marcello}
\email{marcello@gravity.phys.nagoya-u.ac.jp}

\date{\today}%

\begin{abstract}
We derive the functional Schr\"odinger equation for quantum fields in curved spacetime in the semiclassical limit of quantum geometrodynamics with a Gaussian incoherent dust acting as a clock field. We perform the semiclassical limit using a WKB-type expansion of the wave functional in powers of the squared Planck mass. The functional Schr\"odinger equation that we obtain exhibits a functional time derivative that completes the usual definition of WKB time for curved spacetime, and the usual Schr\"odinger-type evolution is recovered in Minkowski spacetime.
\end{abstract}

\maketitle



\section{Introduction}

The Hamiltonian dynamics of quantum fields on a classical curved spacetime of the kind envisioned by Tomonaga and Schwinger \cite{TOM46,SCH48} can be recovered as a WKB-like approximation of canonical quantum gravity \cite{DEW67,KIE09} in the form of a functional Schr\"odinger equation (we work with units where $c = \hbar = 1$):

\eq{
	i \int_\Sigma d^3x G_{abcd} \pfrac{S_0}{\gamma_{ab}}\pfrac{\chi}{\gamma_{cd}} := i \int_\Sigma \pfrac{\chi}{\tau} = H_\phi \chi \label{eq:WKBT}
}

\noindent for the wave functional $\chi$ of matter fields governed by the Hamiltonian $H_\phi$ on the spatial hypersurface $\Sigma$ \cite{KIE94,KIE18,CHA19} (see \cite{KTV18} for a recent comparison of this approach to the Born--Oppenheimer method). Here, $\gamma_{ab}$ are the components of the spatial metric, $G_{abcd}$ is the DeWitt super-metric, and $S_0$ is a solution to the Hamilton--Jacobi--Einstein equation that describes the background geometry. However, the local ``many-fingered'' WKB time $\tau[\gamma_{ab}(x)]$ (a functional of the spatial metric) that emerges from said approximation is identified \textit{ad hoc} with that appearing in the Tomonaga--Schwinger equation, and it most remarkably fails for the simplest case: the flat Minkowski spacetime (in which case $S_0$ is a constant).

When we work with quantum matter fields in flat spacetime, we do so by implicitly assuming that the gravitational effect of the quantum fields' energy distribution on the background can be neglected. If we do so, however, the Schr\"odinger-type evolution cannot be obtained as a semiclassical approximation of canonical quantum gravity: If we assume a truly flat background, WKB time disappears from the Schr\"odinger equation, although the latter is usually taken for granted as a special case of the more generic evolution in curved spacetime. Even though this inconsistency sheds a shadow also on the validity of the definition of WKB time for generic curved spacetimes, it seems to have drawn little to no attention in the literature (see, e.g., at the very end of chapter 2 in \cite{KIE94} and section 3 of \cite{KIE94b}).

In this paper, we would like to draw attention to this issue and provide one viable solution that can be applied to generic curved spacetimes. We do so by considering the semiclassical approximation of quantum geometrodynamics where, from the beginning, we introduce a physical reference clock field in the form of a Gaussian reference dust \cite{KUC90}. This might be pictured as a straightforward quantum realization of the set of free-falling oscillators thinly dispersed in spacetime envisioned by Einstein. The Gaussian time condition is imposed before variation of the action, and time reparametrization invariance is then recovered by parametrizing the new action and promoting Gaussian time to a scalar field labeled by new free spacetime coordinates. The semiclassical limit of the quantized theory results in a functional Schr\"odinger equation for the quantum state of matter fields, as observed on the hypersurfaces determined by the clock field. The functional time derivative operator in \eqref{eq:WKBT} is, in this case, naturally completed with a functional derivative with respect to the clock field configuration 
When we adopt a Gaussian reference frame, the operator is reduced, very intuitively, to a local ``material derivative'' that takes into account both the intrinsic time-dependence of the matter quantum state inherited by the clock as well as the evolution on the background ``medium'': spatial geometry.

This paper is structured as follows: In Section \ref{1}, we will briefly review the elements of the canonical description of the Gaussian reference fluid in the context of quantum geometrodynamics to the extent relevant to our discussion; in Section \ref{2}, we will expand the phase of the total wave functional that solves the quantized Hamiltonian constraint in inverse powers of the Planck mass squared, thus obtaining the Hamilton--Jacobi--Einstein equation describing the background geometry and, at the higher order, the functional Schr\"odinger equation for matter-plus-clock fields; in the same section, we describe the case of the Gaussian reference frame and the explicit recovery of a Schr\"odinger-type evolution in Minkowski spacetime. A discussion of our result with relevant conclusions is presented in Section \ref{3}.

\section{Canonical Reference Dust}
\label{1}

As we are only interested in the Gaussian time condition and keep the spatial frame unaffected by reparametrization, we will closely follow a simplified version of \cite{KUC90}, which we recommend to the reader for its in-depth analysis of the various issues in the classical and quantum regime related to employing a phenomenological fluid to implement coordinate conditions and address the definition of geometrodynamical observables. For a recent study of the perturbation theory of Gaussian dust in an FLRW cosmological scenario, see \cite{GHLS20}.

In the Lagrangian formalism of general relativity, the Einstein field equations can be derived from the Einstein--Hilbert action

\eq{
	S^G[g_{\mu\nu}] = \int_\mM d^4x \sqrt{-g} R[g_{\mu\nu}]  \label{eq:SEH}
}

\noindent by variation of the spacetime metric $g_{\mu\nu}(x)$ of the unbounded manifold $\mM$. We take a vanishing cosmological constant for simplicity and do not yet explicitly introduce the gravitational scale before the action in order to simplify notation.

The diffeomorphism invariance of the theory allows for any choice of spacetime coordinates. Consider the Gaussian time condition

\begin{equation}
\begin{aligned}
& g^{00} + 1 = 0 \label{eq:tc} \, ,
\end{aligned}
\end{equation}

\noindent which fixes as a constant everywhere the normal proper time separation between hypersurfaces of constant times. This choice can be imposed before variating the action by introducing the coordinate time condition \eqref{eq:tc} through a Lagrange multiplier $L = L(x)$ at the level of the action \eqref{eq:SEH}. This results in an extra term

\eq{
	S^D[g_{\mu\nu}, L] := - \frac{1}{2} \int_\mM d^4x \sqrt{-g} \,  L \left( g^{00} + 1 \right)\, .
}

The broken time reparametrization invariance can be restored by parametrization of the action: We promote the Gaussian time to a variable $t \to T(x)$ labeled by new arbitrary coordinates $x$. The new action must be invariant under the transformation of the new coordinates and must be reduced to the old action when Gaussian time is adopted, which leads to

\eq{
	S^D[g_{\mu\nu}, L, T] = -\frac{1}{2} \int_\mM d^4x L \left( 1 + g^{\mu\nu} T_{,\mu}T_{,\nu}\right) \, . \label{eq:SGF_X}
	}

Variation of \eqref{eq:SGF_X} with respect to $g_{\mu\nu}$ provides the vacuum Einstein field equations obtained from \eqref{eq:SEH} with the stress--energy--momentum tensor 

\eq{
	\mathcal{T}^{\mu\nu} = L U^\mu U^\nu \, ,
}

\noindent where

\eq{
	U^\mu :=  - g^{\mu\nu} T_{,\nu} \, 
}

\noindent is the four-velocity of the source. Variation with respect to $T$ gives the dynamical equation 

\eq{
	(L U^\mu)_{;\mu} = 0 \, ,
}

\noindent which describes the source as an incoherent dust. Assuming $L > 0$, the weak, strong, and dominant energy conditions are satisfied, and $T(x)$ constitutes a good candidate as a physical clock field.

In the ADM formalism \cite{ADM}, one performs a $3+1$ decomposition of the metric

\begin{equation}
\begin{aligned}
	g_{00} & = N_aN^a - N^2 \\
	g_{0a} & = N_a \\
	g_{ab} & = \gamma_{ab} \label{eq:ADM},
\end{aligned}
\end{equation}

\noindent where $N$ is the lapse function, $N^i$ is the shift vector, and $\gamma_{ab}$ with $a,b \in (1,2,3)$ is the induced spatial metric. Using these new variables, spacetime is described by the time propagation of the three-dimensional hypersurfaces of constant time corresponding to the chosen arbitrary foliation. The parametrized action takes the form

\eq{
	S_{ADM} = \int_\mathbb{R} dt \int_\Sigma d^3x \left( \pi^{ab} \dot{\gamma}_{ab} + P \dot{T} - N H -N^a H_a )\right), \label{eq:SADM}
}

\noindent where $P$ is the canonical momentum conjugate to $T$. Variation with respect to $N$ and $N^i$ gives the constrained Hamiltonian and momenta

\begin{equation}
\begin{aligned}
	& H  := H^G + H^D = 0  \\
	& H_a := H^G_a + H^D_a = 0 \, , \label{eq:C}
\end{aligned}
\end{equation}

\noindent where the constrained Hamiltonian and momenta for gravity are

\begin{equation}
\begin{aligned}
	& H^G =  \gamma^{-\frac{1}{2}} \left( \pi_{ab} \pi^{ab} - \frac{1}{2} (\pi^a_a)^2\right) - \gamma^\frac{1}{2} {}^{(3)}R[\gamma_{ab}] \\
	& H^G_a = - 2 {\pi_a^b}_{;b}\, , \label{eq:HG}
\end{aligned}
\end{equation}

\noindent and for the clock: 

\begin{equation}
\begin{aligned}
	& H^D := n P \\
	& H^D_{i} := P T_{,i} \, , \label{eq:HF}
\end{aligned}
\end{equation}

\noindent where $n = (1 + \gamma_{ab} T^{,a}T^{,b} )^\frac{1}{2}$.

The equations of motion for $\gamma_{ab}$ and $\pi^{ab}$ can be obtained from the super-Hamiltonian of the canonical action \eqref{eq:SADM}:

\eq{
	\bar{H} = \int_\Sigma d^3x \left( N H + N^a H_a \right) \, . \label{eq:HADM}
}

We quantize the theory by promoting the canonical variables to operators and requiring the physical state to be annihilated by the constraint operators

\eq{
	& \widehat{H} \Psi = 0 \label{eq:QHC} \\
	& \widehat{H}_i \Psi = 0 \, . \label{eq:QMC}
}

In the next section, we will focus on the Hamiltonian constraint \eqref{eq:QHC},  which gives way to a Schr\"odinger-type evolution in the $\{T(x)\}$-representation. This evolution is consistent and unambiguous with respect to the foliation choice only if the commutators of the constraints vanish, a condition that originates from the quantization of the classical equations of motion. This condition depends, in turn, on the factor ordering of the geometrodynamical operators. As the focus of the present work concerns only the emergence of time in the semiclassical limit, we will not deal with the open issues of quantum geometrodynamics as long as they do not affect the general validity of the semiclassical approach. In the specific case, the choice of operator ordering results in non-derivative terms in the functional Schr\"odinger equation for matter fields, and since this does not affect the definition of semiclassical time, we will adopt the trivial ordering.

However, it is worth noticing that, when one quantizes the incoherent dust model, one is not deprived of a consistent definition of the probability density function for $\gamma_{ij}$ on the embedding $T(x)$, as it occurs instead for the more general Gaussian reference fluid where the full Gaussian coordinate conditions are imposed and all four spacetime coordinates are promoted to variables. Consequently, the clock field model may help define a conserved positive inner product in superspace and, although the quantized theory may still present other problems, it may provide a valid starting point to consider them.

\section{Functional Schr\"odinger Equation}
\label{2}

The description of matter fields with non-derivative coupling to gravity can be included straightforwardly in the canonical formalism. We will use a scalar field $\phi$ as their representative. In the $\{\gamma_{ij}(x),T(x), \phi(x)\}$-representation, the Hamiltonian constraint \eqref{eq:QHC} becomes

\eq{
	 i \pfrac{\Psi}{T} = \left( \mH_G \left[ \gamma_{ab}, - i \delta/\delta \gamma_{ab} \right] + \mH_\phi \left[ \phi, - i \delta/\delta \phi, \gamma_{ab} \right] \right) \Psi \label{eq:sch} \, .
}

In this section, we will imply spatial integration throughout this equation and its consequences in order to simplify notation.

To the purpose of discussing the semiclassical limit, we will introduce in \eqref{eq:SEH} the gravitational scale $ M := 1 / 32 \pi G = (M_P/2)^2$, $M_P$ being the reduced Planck mass. The geometrodynamical Hamiltonian density will read

\eq{
	\mH^G := \left( 2 M \sqrt\gamma \right)^{-1} G_{ijkl}\pi_{ij}\pi^{kl} - \left( 2 M \sqrt\gamma \right) \, {}^{(3)}R \, , 
}

\noindent where $G_{ijkl}$ is the index-lowering DeWitt metric of superspace, the configuration space of general relativity.

We perform the semiclassical limit of \eqref{eq:sch}, following for the most part \cite{KIE94}, and consider an expansion in powers of $M$ of the wave functional 

\eq{
	\Psi[ \gamma_{ab}, \phi, T] =  \exp \left[ i M \sum_{n=0}^\infty \frac{1}{M^n} S_n[\gamma_{ab}, \phi, T] \right] \, , \label{eq:EXP}
}

\noindent where $S_n \in \mathbb{C}$ in general.

Substituting in \eqref{eq:sch} and equating terms of equal power, the highest order ($M^2$) contribution comes only from the kinetic term of the matter Hamiltonian, and gives us that the leading term $S_0 = S_0[\gamma_{ab}, T]$ is independent on the matter fields. Although it is not necessary, we will require that $S_0$ is also independent on $T$, as we want to treat gravity as classical and the clock field as quantum.

At the next order ($M^1$), we retrieve the Hamilton--Jacobi--Einstein equation \cite{PER62} with Hamilton's principal function $S := M \, S_0$:

\eq{
	\mH^{G} \left[ \gamma_{ab}, \pfrac{S}{\gamma_{ab}}\right] = 0 \, , \label{eq:hje}
}

\noindent which provides a classical description of the vacuum background space equivalent to the $G_{00}$ component of Einstein's field equations \cite{GER69,MON72}. Equation \eqref{eq:hje}, which is integrated over the space coordinates, holds true for any value of coordinate time. Rather than considering the evolution between spatial hypersurfaces that takes place in spacetime \cite{KUC92}, we picture the evolution to take place in the ''super-spacetime" $\mathcal{S}\times\mathcal{T}$ given by the cartesian product of the superspace $\mathcal{S}$ spanned by the functions $\gamma_{ab}(x)$ and the configuration space $\mathcal{T}$ of the clock field $T(x)$.

Proceeding with the semiclassical expansion, at order $M^0$, we have

\eq{	
\left[ -i \pfrac{S_1}{T} + G_{abcd} \left( \pfrac{S_0}{\gamma_{ab}}\pfrac{S_1}{\gamma_{cd}} - \frac{i}{2} \frac{\delta^2 S_0}{\delta \gamma_{ab} \delta \gamma_{cd}} \right) +  \mH_\phi \right] e^{i\left( M S_0 + S_1 \right) }  = 0 \label{eq:M0} \, .
}

We decompose $S_1$ into a pure complex part only dependent on geometry alone and a complex part $S_1''$ dependent also on the field configurations

\eq{
	S_1[\phi, T | \gamma_{ab}] =  i S_1'[{\gamma}_{ab}] + S_1''[\phi, T | \gamma_{ab}] \quad S_1' \in \mathbb{R} \, , \quad S_1'' \in \mathbb{C} \, ,
}

\noindent and define the matter wave functional

\eq{
	\chi[\phi, T|\gamma_{ab}] := \exp \left( i S_1'' [\phi, T | \gamma_{ab}] \right) \, ,\label{eq:chi}
}

\noindent where we use the vertical bar in the argument to distinguish the dynamical degrees of freedom of the functional from the other functions of spacetime. At this order of approximation, the total wave functional can be written as

\eq{
	\Psi \approx \left( \varrho[{\gamma}_{ab}]^\frac{1}{2} e^{i \tilde{S}[\bs{\gamma}]  } \right) \, \chi[\phi, T|{\gamma}_{ab}] := \psi[{\gamma}_{ab}] \, \chi[\phi, T | {\gamma}_{ab}],
}

\noindent where we have defined

\eq{
	\varrho[{\gamma}_{ab}] := \exp \left( -2 S_1'[{\gamma}_{ab}] \right) \, .
}

Let us then impose the conservation law

\eq{
	G_{abcd} \pfrac{ j^{cd}}{\gamma_{ab}} = 0 \label{eq:c}
}

\noindent for the current density

\eq{
	j^{ab} := \frac{1}{M} \, \varrho \, \pfrac{S}{\gamma_{ab}} \, . \label{eq:j}
}

Equation \eqref{eq:M0} then gives
\eq{
	\int_\Sigma d^3x \, \left( \frac{1}{i}\frac{D \,}{D\, T} + \mH_\phi \right) \chi[\phi, T| \gamma_{ab}] = 0  \, \label{eq:ST1} \, ,
}
where we have reinserted the spatial integration for the sake of clarity, and we have defined the functional operator
\eq{
	\frac{D}{D \, T} := \pfrac{}{T} + u_{ab} \pfrac{}{\gamma_{ab}} \label{eq:DT} \, ,
}
where
\eq{
	u_{ab} =  G_{abcd} \pfrac{S_0}{\gamma_{cd}} 
}
are the geometrodynamical ``velocities'' $u_{ab}$ that can be obtained in the Hamilton--Jacobi approach from the variation of Hamilton's principal function $S$ governing \eqref{eq:hje}.

Equation \eqref{eq:ST1} is valid for generic spacetimes, and it does not break diffeomorphism invariance, as $T(x)$ simply determines the hypersurface on which we decide to observe or register the matter state. In this picture, the time evolution of the matter field $\phi(x)$ is implicitly determined by its correlations with the values of the clock field $T(x)$, which one can physically measure in principle. We then have an application at the semiclassical level of the conditional probability interpretation proposed by Page and Wootters \cite{PAG83}, where the matter state $\chi[\phi| \gamma_{ab}, T']$ at a given ``time'' $T'(x)$ is obtained by conditioning the total ``timeless'' wave function $\chi[\phi, T | \gamma_{ab}]$ with the state of the clock field in the configuration $T'(x)$.

As a special case, we may choose our coordinates so that the value of the clock field $T(x)$ is constant on hypersurfaces of constant (Gaussian) time, $T(x) = t$. Then, the clock field will appear in the matter wave functional as a simple parameter rather than a physical variable. In this case, the structure of the augmented superspace $\mathcal{S} \times \mathcal{T}$ becomes redundant, and we can simply work on the superspace spanned by the Gaussian metric field $\gamma_{ab}(x)$
\eq{
	( \gamma_{ab}(t, \bs{x}), T(t,\bs{x}) = t ) \to \gamma_{ab}(x) \, .
}
%

The operator on the left-hand side of \eqref{eq:ST1} reduces to
\eq{
	\frac{D}{D \, T} \quad \to \quad \frac{D}{D \, t} = \frac{\pa}{\pa t} + \dot{\gamma}_{ab} \pfrac{}{\gamma_{ab}} \, , \label{eq:DT2}
}
where the velocities $u_{ab} = \dot{\gamma}_{ab}$ are consistent with the equations of motion given by the super-Hamiltonian \eqref{eq:HADM}
\eq{
	\dot{\gamma}_{ab} = N  G_{abcd} \pfrac{S_0}{\gamma_{cd}} + N_{a;b} + N_{b;a}\, \label{eq:v2}
}
for the case of Gaussian coordinates $(N =1, N_a = 0)$.

In analogy with classical continuous mechanics, the ``material derivative'' \eqref{eq:DT2} takes into account both the intrinsic time evolution of the state (i.e., its dependence on clock time) and the evolution of the background ``medium'' (the spatial metric field). When one performs the semiclassical limit starting from the usual Wheeler--DeWitt equation without clock field, only the second (``convective'') part of \eqref{eq:DT2} is present, and thus, it cannot account for the time evolution in static spacetime.

The general solution to \eqref{eq:ST1} at Gaussian time $t$ can be expressed in terms of any solution at time $t_0<t$ as
\eq{
	\chi[\phi | \gamma_{ab}, t] = \exp \left[ - i \int d^3x \int_{\bs{\gamma}} \, \mH_\phi \, ds\right] \chi[\phi | \gamma_{ab}, t_0] \, ,
}
where the line integral is taken along the classical trajectory $\bs{\gamma}(x)$ that solves the Hamilton--Jacobi--Einstein equation. The standard evolution in Minkowski spacetime is thus retrieved, together with that of any spacetime described by a Gaussian foliation. For more general coordinates, we must rely on Equation \eqref{eq:ST1}.

\section{Discussion and Conclusions}
\label{3}

In this work, we have shown how  the introduction of a dynamical Gaussian clock field can provide a semiclassical limit for canonical quantum gravity where matter fields are described by a Schr\"odinger-type evolution with a notion of time that also remains valid when the geometrodynamical momentum vanishes, such as in Minkowski spacetime. The semiclassical limit consists of a classical gravitational background that is still consistent with the ``timeless'' equations of motion of general relativity in the form of the Hamilton--Jacobi--Einstein equation, and quantum matter fields that evolve according to a functional Schr\"odinger equation where the usual WKB time derivative is extended into a functional derivative operator that takes into account the dependence of the matter state on the configuration of the clock field. For Gaussian time, this operator is reduced to the analogue of a material derivative where both the explicit dependence on clock time and the time dependence of the classical gravitational background are taken into account.

While the Gaussian clock field has been our tool in order to recover a viable time evolution for quantum fields in curved (and flat) spacetime, various alternative derivations of Schr\"odinger-type evolutions for quantum gravity have been proposed since the first formulation of the canonical quantization of the theory, especially in relation to the so-called problem of time (for an extensive review that also includes the Page and Wootters approach cited above, see \cite{AND17}). Other models may equally provide, in a semiclassical approach, viable solutions to the problem addressed here. Our primary concern has been the recovery of an appropriate semiclassical limit for the dynamics of quantum matter fields, and we do not claim that it provides a viable definition of time in the quantum gravity regime, nor that such a notion is necessary to begin with.

In a generic curved spacetime, the Hamiltonian formulation of quantum field theory is problematic \cite{FUL89}, and it has been questioned whether the Tomonaga--Schwinger equation is indeed able to consistently describe the unitary time evolution of quantum fields between spacelike hypersurfaces (see, e.g., \cite{TOR99,COL11}). Our treatment, which results in a functional Schr\"odinger-type equation \eqref{eq:ST1}, the solution of which describes the correlation of the configurations of matter fields with the configuration of the clock without any explicit choice of foliation, may help to address this issue.

Incidentally, we would like to finally draw attention to the fact that, unlike in our case, the Tomonaga--Schwinger equation was introduced in the interaction picture. This detail does not seem to have received much attention either, although it might have some relevance to the interpretation and regularization of the ``wave function of the universe'' that solves the Wheeler--DeWitt equation.

\vspace{6pt}




\end{document}